\documentclass[]{aastex631}

\usepackage{multirow}
\usepackage{booktabs}
\usepackage{amsmath}
\usepackage{subfigure}

\begin{document}

\title{Modeling of Nitric Oxide Infrared radiative flux in lower thermosphere: a machine learning perspective}

\author{Dayakrishna Nailwal}
\affiliation{Indian Institute of Technology Roorkee, Roorkee-247667, Haridwar,India}

\author{M V Sunil Krishna}
\altaffiliation{Correspondence to: M V Sunil Krishna, mv.sunilkrishna@ph.iitr.ac.in}
\affiliation{Indian Institute of Technology Roorkee, Roorkee-247667, Haridwar,India}

\author{Alok Kumar Ranjan}
\affiliation{Indian Institute of Technology Roorkee, Roorkee-247667, Haridwar,India}

\author{Jia Yue}
\affiliation{NASA's Goddard Space Flight Center, 8800 Greenbelt Rd, Greenbelt, 20771, USA}
\affiliation{Catholic University of America, 620 Michigan Avenue, NE, Washington DC, 20064, USA}

\begin{abstract}
Nitric Oxide (NO) significantly impacts energy distribution and chemical processes in the mesosphere and lower thermosphere (MLT). During geomagnetic storms, a substantial influx of energy in the thermosphere leads to an increase in NO infrared emissions. Accurately predicting the radiative flux of Nitric Oxide is crucial for understanding the thermospheric energy budget, particularly during extreme space weather events. With advancements in computational techniques, machine learning (ML) has become a highly effective tool for space weather forecasting. This effort becomes even more worthwhile considering the availability of two decades of continuous NO infrared emissions measurement by TIMED/SABER along with several other key thermospheric variables. We present the scheme of development of an ML-based predictive model for Nitric Oxide Infrared Radiative Flux (NOIRF). Various ML algorithms have been tested for better predictive ability, and an optimized model (NOEMLM) has been developed for the study of NOIRF. This model is able to extract the underlying relationships between the input features and effectively predict the NOIRF. The NOEMLM predictions have very good agreements with SABER observation during quiet time as well as geomagnetic storms. In comparison with the existing TIEGCM model, NOEMLM has very good performance, especially during extreme space weather conditions. The results of this study suggest that utilizing geomagnetic and space weather indices with ML/AI can serve as superior parameters for studying the upper atmosphere, as compared to focusing on specific species having complex chemical processes and associated uncertainties in constituents. ML techniques can effectively carry out the analysis with greater ease than traditional chemical studies.
\end{abstract}

\section{Introduction}
The upper atmosphere of the Earth (mesosphere and thermosphere) is continuously affected by solar radiation, geomagnetic storms, and other space weather events like solar flares, Solar Energetic Particle (SEPs), Solar Proton Events (SPEs), etc \citep{kutiev2013solar,sinnhuber2012energetic,lei2008rotating}. Direct energy deposition in the MLT region in the form of X-rays and Extreme Ultra-Violet (EUV) radiation triggers various photochemical reactions. However, the interaction of highly energetic solar wind plasma associated with the frozen-in magnetic field to the Earth’s magnetic field leads to the deposition of a part of the solar wind energy into the upper atmosphere in the form of energetic particle precipitation and Joule heating. This energy subsequently results in enhanced auroral activity. The electric field and energy input through this interaction causes perturbations in temperature, densities of neutral and charged species, ionospheric electric fields, and dynamics in the MLT region \citep{fuller1994response,lu1998global,knipp1998overview,richmond2000upper,knipp2004direct,turner2009geoefficiency,lu2016high}.

The MLT region is mainly composed of the primary atmospheric gas species such as nitrogen (N$_2$) and oxygen (O, O$_2$) as well as trace gases like carbon dioxide (CO$_2$), ozone (O$_3$), various oxides of nitrogen (NOx), helium (He) and neon (Ne). These trace gases significantly contribute to the regulation of long-term atmospheric compositional changes and also play a crucial role in climate change. Therefore, there is an overarching need to quantify these trace gases and understand their mechanistic descriptions so that we can make better climate change predictions. The role of greenhouse gases (CO$_2$, N$_2$O, and O$_3$) in climate change is well understood. In addition to these species, there are several such other trace gases having the potential to change long-term climatic conditions (methane (CH$_4$), oxides of nitrogen (NO, N$_2$O, NO$_2$), etc.) \citep{ramanathan1985trace,hansen2007climate}.
NO is a very important trace species in the MLT region and plays an essential role in thermospheric chemistry and energy budget. Due to its low ionization energy, NO is highly reactive and acts as a terminal ion in many of the charge exchange and ion-molecule/atom chemical reactions in the thermosphere \citep{barth1992nitric}. The formation of NO in MLT is prominently enhanced by the incoming energetic particle flux during active auroral periods \citep{bailey2002model,baker2001relationships}. The formation mechanism of NO in the thermosphere is given in the chemical scheme from R1 to R8 \citep{bharti2018storm,ballard1993observations, rusch1991diurnal,solomon1982photochemical}.

\begin{center}

N$_2$ + $h\nu$  \ \ \  $\rightarrow$ \ \ \  N($^4$S) + N($^4$S)     \hspace{3.55cm}        (R1-a)\\
N$_2$ + $h\nu$   \ \ \  $\rightarrow$ \ \ \ N($^4$S) + N($^2$D)          \hspace{3.5cm}         (R1-b)\\
N$_2$ + $h\nu$   \ \ \  $\rightarrow$ \ \ \  N($^4$S) + N($^2$P)          \hspace{3.55cm}       (R1-c)\\
$e^*$ + N$_2$     \ \ \  $\rightarrow$ \ \ \ N($^2$D) + N($^4$S) + 
 $e^*$   \hspace{3.1cm}        (R2)\\
N($^2$D) + O$_2$   \ \ \  $\rightarrow$ \ \ \ NO + O      \hspace{4.3cm}        (R3)\\
N($^4$S) + O$_2$   \ \ \  $\rightarrow$ \ \ \ NO + O                \hspace{4.4cm}      (R4)\\
NO($\nu$=0) + O   \ \ \  $\rightarrow$ \ \ \ NO($\nu$=1) + O    \hspace{3.1cm}        (R5)\\
NO($\nu$=1)   \ \ \  $\rightarrow$ \ \ \ NO($\nu$=0) + $h\nu$ (5.3 $\mu m$)  \hspace{2.3cm}   (R6)\\
NO + N($^4$S)  \ \ \  $\rightarrow$ \ \ \ N$_2$ + O   \hspace{4.4cm}    (R7)\\
NO($\nu$=1) + O   \ \ \  $\rightarrow$ \ \ \  NO + O  + Kinetic energy  \hspace{1.35cm}    (R8)\\
\end{center}
Where $h\nu$ represents the photon energy, $e^*$ represents an energetic electron, N($^4$S), N($^2$P) and N($^2$D) are the excited states of atomic nitrogen, $\nu$=0 and $\nu$=1 are the vibrational ground and first excited states of NO.

At low and mid-latitudes, primarily the energy input from the sun in the form of soft X-rays and Extreme Ultra-Violet (EUV) radiation dissociates the Nitrogen (N$_2$) molecules (represented by (R1-a), (R1-b), (R1-c), which combine with Oxygen (R3, R4) and form NO \citep{richards1981photodissociation,bailey2002model,barth2003global}. 
At higher latitudes, the energy inputs from particle precipitation due to auroral and geomagnetic activity lead to the formation of NO by dissociating nitrogen (N$_2$), as shown in R2. This particle precipitation also contributes to the enhanced temperature in mesospheric and thermospheric altitudes. The enhanced temperature, in combination with the available oxygen density, leads to the production of vibrationally excited NO ($\nu$=1) by collisional excitation (shown by R3 \& R4) \citep{kockarts1980nitric,rusch1991diurnal, sharma1996production}. 
In addition, the joule heating and particle heating during a geomagnetic storm also contribute to the large-scale production of NO in polar latitudes by being a major source for the upwelling of N$_2$ rich air and elevating the temperature. It is important to note that the processes shown in (R3) and (R4) are the dominant sources of NO near 110 km and above 130 km, respectively. The vibrationally active NO can also be quenched by the process shown in R7 overall latitude regions \citep{richards2004increases,siskind1989response}.

The NO is a vibrationally active molecule, and due to the presence of the 5.3 $\mu$m  infrared (IR) active vibrational mode, it plays a vital role in regulating the thermospheric energy budget \citep{lu2010relationship}. The rate of collisional excitation of NO by atomic oxygen into the vibrationally excited state is given by [k$_1$[O] $exp(-2700/T$) sec$^{-1}$], where k$_1$ is the collisional quenching rate coefficient of NO ($\nu$=1) by atomic oxygen (k$_1$ = 4.2 $\times$ 10$^{-11}$cm$^{-3}$ sec$^{-1}$) as reported by \cite{hwang2003vibrational}. 
This collisional excitation rate of NO in the MLT region strongly depends on the temperature and the concentration O \citep{sharma1998model,ranjan2023aspects}. Thus, the infrared flux of 5.3 $\mu$m emission is highly sensitive to changes in temperature, [O] and [NO]. The enhanced energy deposition in the lower thermosphere during geomagnetic storms leads to an increase in temperature and enhances the production of NO (R3, R4). Due to increased temperature, the collisional excitation of NO($\nu$=0) with atomic oxygen (O) populates more NO in its excited vibrational state NO($\nu$=1) (R5). The de-excitation of NO (R6) leads to the emission of IR radiation at 5.3 µm into the space. A higher population of NO in the excited state ($\nu$=1) results in enhanced IR emission of 5.3 µm during the geomagnetic activity and particle precipitation events. The vibrationally active NO can also be quenched by the process shown in R7 \& R8 overall latitude regions \citep{richards2004increases, siskind1989response}. A large part of energy input during intense space weather events into the MLT region is captured by the NO, and it is re-radiated into space by means of 5.3 $\mu$m infrared flux (IRF). Hence, the NO acts as an effective cooling agent to control the overall energy and temperature of the MLT region as a natural thermostat \citep{mlynczak2005energy}.

During the geomagnetic storm of April 2002, the zonal mean emission by NO at 5.3 µm, as observed by Sounding of the Atmosphere using Broadband Emission Radiometry (SABER), suggested an increase of nearly 5-7 times in the total IRF \citep{mlynczak2003natural}. Several other studies \citep{ranjan2023aspects,bharti2018storm, lu2010relationship,mlynczak2010short} have confirmed the role of NO 5.3 µm infrared flux (IRF) in effectively cooling the upper atmosphere during extreme space weather events. Further, an investigation of geomagnetic storms using SABER measurements and TIEGCM simulations found that under the disturbed condition, about 80\% of energy input into the thermosphere is released back into space through NO 5.3 µm infrared emission \citep{mlynczak2003natural,mlynczak2005energy,lu2010relationship}.
Thus, the NO species can be a very good indicator for sensing the changes in the state, structure, and dynamics of MLT regions during intense geomagnetic storms and particle precipitation events. Under typical sunlight conditions in the MLT region, the chemical lifetime of NO is approximately one day. However, during polar winter nights when sunlight is absent, NO may have a lifetime of several weeks since there is no sunlight available to initiate photodissociation, and there are no other significant processes to destroy the NO \citep{bailey2022sounding}. Due to this prolonged lifetime in polar regions during the winter nights, NO has sufficient time to descend into the stratosphere without undergoing photochemical destruction. As NO descends near the polar region, it is partially converted into nitrogen dioxide (NO$_2$) and finally reaches into the stratosphere in the form of NO and NO$_2$, where they react with ozone and destroy it \citep{perot2014unusually,solomon1982photochemical}.

A significant knowledge about NO and its variability in the MLT region is based on satellite-based observations. The satellite experiments Solar Mesosphere Explorer (SME) \citep{barth1988solar} and Halogen Occultation Experiment (HALOE) \citep{russell1993halogen} have shown that NO in the thermosphere is highly variable with altitude and time. 
The observations of NO volume emission rate (NOVER) by SABER and NO abundance by Student Nitric Oxide Explorer (SNOE) have found that the radiance and abundance of NO in the thermosphere are strongly correlated with solar irradiance and geomagnetic activities \citep{barth2003global}. The emission by NO is highly susceptible to changes caused by solar and geomagnetic storms and has been reported by \citep{bharti2018storm, bharti2019radiative,tang2017global,knipp2017thermospheric, knipp2013thermospheric,li2018comparison,li2019eofs, mlynczak2010short,mlynczak2003natural,mlynczak2005energy}.

\begin{figure}
\centering
\includegraphics[scale=0.75]{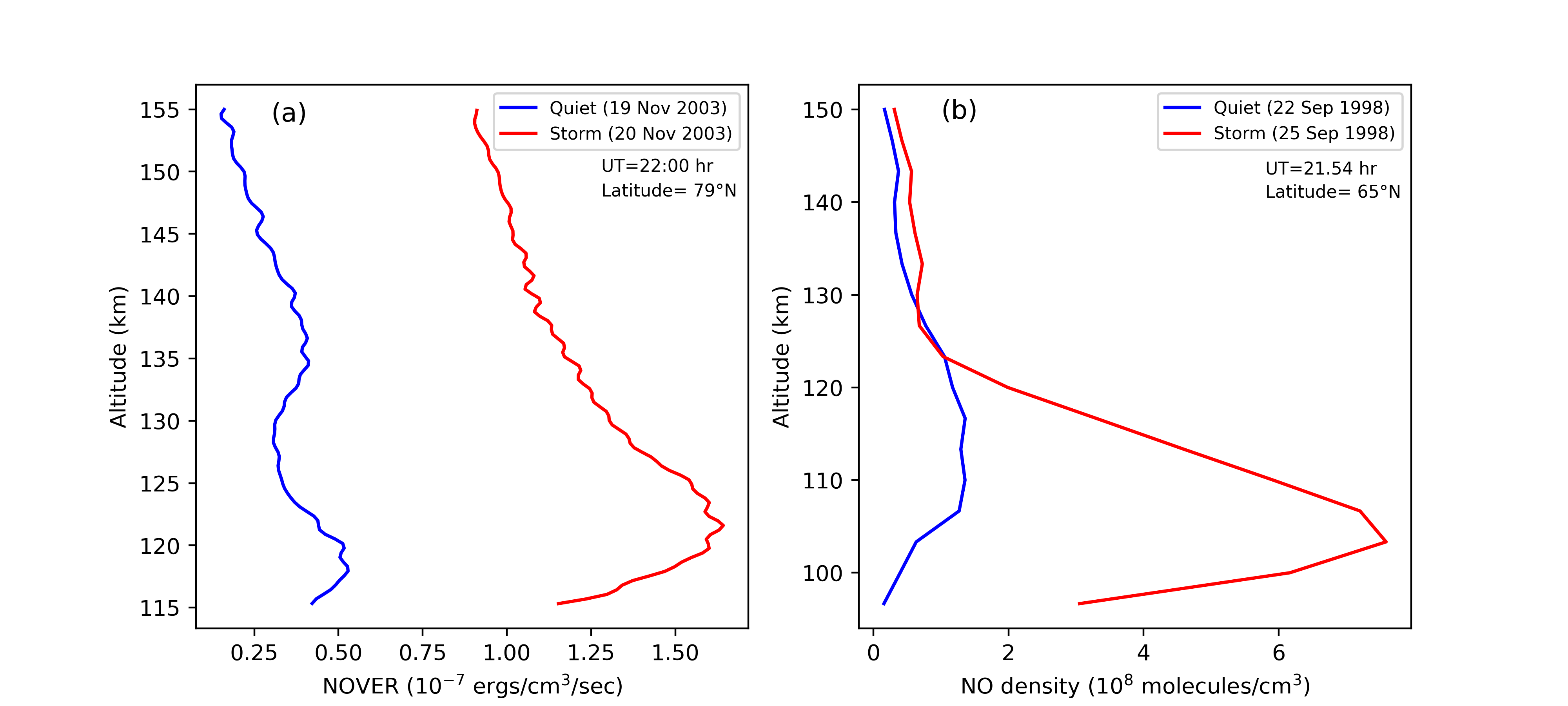}
\caption{Variation of (a) NO Volume Emission Rate, and (b) NO abundance as measured by SABER and SNOE during quiet and storm time.}
\end{figure}

Figure 1 (a) shows the variability of NO infrared flux at 5.3 µm between a quiet and disturbed day as observed by SABER. It can be seen that the infrared flux is enhanced by an order of magnitude during an intense space weather event in comparison to a quiet control day. This enhancement in the flux is a result of a combined rise in temperature, NO density, and its collision through atomic oxygen, as discussed earlier. The NO abundance measured from SNOE, as shown in Figure 1(b), also supports the role of intense space weather events in increased NO production. Thus, the role of NO in balancing the energetics of the mesosphere-thermosphere is very well established. 

Understanding NO variation in the thermosphere becomes a key feature of knowing the thermospheric temperature structure and energy balance. The enhancement and transport of nitric oxide is part of the global effect that is generally set up during intense space weather events in terms of large-scale traveling ionospheric disturbances from the polar region towards the lower latitudes. In order to completely understand the space weather modulation of Earth’s atmosphere, NO can be a key parameter. Continuous and simultaneous global measurements of infrared flux and abundance can help us comprehensively understand space weather effects on Earth’s atmosphere.  However, due to various limitations (limited orbital geographic sampling and discontinuous coverage of specific satellite observations), it becomes difficult, expensive, or sometimes impossible to achieve simultaneous global coverage, even with the help of advanced observational devices. 
The orbit's limitation on geographic sampling prevents simultaneous real-time global observations. The scope of our understanding is only limited to the latitude, longitude, and time sectors where observations are made. The short-time fluctuations and their effects on the atmosphere are also not captured due to the limited orbital trajectory of the satellites. It will be a worthwhile effort to fill this gap by developing advanced mathematical/computational models that can effectively use the existing data. 
The information on NO abundance and its change during a geomagnetic storm is very important to decipher the chemistry, dynamics, and transport of the MLT region. There have been very limited direct observations of NO in recent times. To estimate the abundance of NO, especially during storm time, various theoretical models have been developed \citep{emmert2022nrlmsis,marsh2004empirical,kiviranta2018empirical}.
However, in theoretical models, the reaction rate coefficients of NO and other constituents have considerable uncertainties. The reaction rate coefficients for NO in the thermosphere have been investigated theoretically and experimentally by \citep{herron1999evaluated,duff2003rate}, but they could not validate the results completely. There have been reports of discrepancies with the modeled NO emission in comparison to the measurements of SABER \citep{chen2018numerical, li2018comparison, sheng2017thermospheric, qian2010model}.

The complex chemical processes and mutual coupling of parameters affected by the vertical and meridional transport play an essential role in the NO emissivity from the upper atmosphere. These complex chemical reactions, along with the associated uncertainties in constituents, make it challenging to identify the relative importance of various reactions and subsequent estimation of reaction rates. 
In recent years, machine-learning modeling techniques have become effective tools in atmospheric and space science due to the availability of large amounts of satellite data volume, powerful computing techniques, and enhanced data processing machines \citep{ball2010data,abadi2016tensorflow,camporeale2018machine}.  The advanced ML/AI models are based on the measurement of the data from the last several years and have a distinct advantage over conventional modeling methods such as theoretical or photochemical models. These models can utilize the vast data sets of multiple modes of measurement and build a unified understanding of the processes. Predictive models have the ability to establish even very weak correlations between several measured parameters against the independent variables and vice versa. The advanced machine learning methods exploiting SABER measurements can provide us with a unique opportunity to estimate and predict the NO infrared flux variation as a function of latitude, longitude, and time. A continuous and accurate prediction of enhanced NOIRF in the thermosphere with respect to latitude, longitude, universal time, and local time during different kinds of geomagnetic activities will definitely improve our understanding of thermospheric energy budget, and the associated thermospheric density response \citep{chen2018numerical,sheng2017thermospheric}.

In this paper, we present a machine learning (ML) approach to understanding the effect of space weather on thermospheric NO infrared emission and predicting the NOIRF. As we are heading into the active phase of solar cycle 25, this model can be a very valuable tool for the prediction of space weather fluctuations and their effect on thermospheric chemistry, dynamics, and energetics. A combined effect of various modes of thermospheric perturbations has a strong signature on NOIRF, and targeting the NOIRF via very sensitive information about the overall correlation of several key variables can be obtained using a machine learning approach. The SABER instrument dataset provides the measurements of NO IRF for the past two decades, spanning from 2002 to date. In terms of the solar cycle, the measurements hold the information on thermospheric variability from the solar cycle 23 (descending phase) and the complete solar cycle 24. It is very important to note that all possible effects of solar cycle variations on the earth’s MLT region are recorded in these measurements. This dataset provides a unique opportunity for us to explore and see the minute correlations by using machine learning models. In this study, we have developed a machine learning-based model for NOIRF primarily based on the SABER dataset.

\section{Data and Model descriptions}
\subsection{SABER Observations}
The dataset of 5.3 µm emission by NO is obtained from the SABER instrument, which is part of NASA’s TIMED (Thermosphere Ionosphere Mesosphere Energetics Dynamics) satellite. SABER provides data for NO IR emission at 5.3 µm. It measures the data in selected sets of wavelengths within the spectral range of 1.27 µm to 17 µm, including 5.3 µm NO emission data. The TIMED satellite was launched into a 625 km near-circular orbit with an inclination of 74 degrees. Every 60 days, orbit precession provides 24-hour local time coverage, while TIMED performs a yaw maneuver to avoid SABER from pointing to the Sun, resulting in few SABER observations around local noon. The SABER observations cover the latitudes range from 52°S to 83°N or from 83°S to 52°N depending on the yaw cycle of the satellite \citep{russell1999overview}.
The SABER instrument has been in operation since 25 January 2002. In this study, the machine learning-based model utilizes the sixteen years of SABER data from 2002 to 2017 as the primary dataset.

\subsection{Model Descriptions}
This paper presents the implementation of machine learning algorithms to predict the NOIRF based on earlier observations. The SABER-measured NO volume emission rate altitude profiles, along with its trajectory, are integrated to calculate the NOIRF. The machine learning model referred to as the Nitric Oxide Emission Machine Learning Model (NOEMLM) is constructed for the target variable NOIRF. The observed values of infrared flux based on current understanding depend on several global variables such as geographic coordinates, geomagnetic indices, solar flux, and local time of measurement. These parameters along the trajectory are obtained from the \href{https://omniweb.gsfc.nasa.gov/}{OMNIWeb} datasets and \href{http://saber.gats-inc.com/data.php}{SABER} measurements \citep{esplin2023sounding}. 
The selection of various features of the machine learning algorithm is entirely based on the current understanding of NO radiative emissions and their variability throughout. Machine learning techniques have the ability to recognize the subtle underlying correlation present between features in addition to the correlation of features with the target variable. The OMNIWeb datasets for the respective conditions are merged with the SABER dataset.

To include all possibilities of correlations between features and target and cross-correlations between features, we have exposed the entire SABER-OMNIWeb dataset corresponding to the observation period of 2002-2014. The selection of this range of data from 2002 to 2014 is based on that this time period includes the descending phase of solar cycle 23 and the ascending phase of solar cycle 24. So, it provides a complete dataset of NOIRF variability during the solar cycle. This dataset can be considered the training dataset. Different machine learning algorithms have been implemented on this dataset, and their performance is evaluated using the ML model output against the actual observations made by SABER from 2015 to 2017. The period from 2015 to 2017 spans the descending phase of solar cycle-24 (2009-2021), which achieved its maximum during 2014. This period includes major solar prominences, which can have a significant impact on the thermosphere radiative cooling like geomagnetic storms, solar flares, and other extreme space weather events. This makes the dataset ideal for the evaluation of NOEMLM’s performance. The target variable (NOIRF) is the integrated NOVER in the altitude range of 115 - 250 km. It is very well established by \citep{mlynczak2003natural,bharti2018storm} that NO IR emission in this altitude region is highly sensitive to the space weather modulation, and Figure 1 (a) also shows the very strong enhancement in NO IR emission during geomagnetically disturbed periods. 

\begin{table*}[h]
\centering
\caption{List of input parameters, their description, unit, and sources.}
\begin{tabular}{|l|l|l|l|}
\hline
Input parameter & Description & Unit & Source\\
\hline
Bz & Z- component of interplanetary magnetic field & nT & OMNIWeb\\
 Dst index &	Disturbance storm time index&	nT&	OMNIWeb\\
 AE index&	Auroral electrojet index&	nT&	OMNIWeb\\
S spot&	Sunspot number&		&OMNIWeb\\
SZA	&Solar zenith angle&	Rad&	SABER\\
Kp&	Geomagnetic planetary K-index&		&SABER\\
F10.7&	Solar radio flux at (10.7 cm)&	SFU	&SABER\\
Ktemp&	Kinetic temperature&	K&	SABER\\
Latitude&	Latitude&	Deg	&SABER\\
Longitude&	Longitude&	Deg & SABER\\
NOIRF&	NO Infrared Radiative Flux&	mW.m$^{-2}$&	SABER\\
\hline
\end{tabular}
\end{table*}

The complete list of features and their respective sources being used for this study are listed in Table 1. In all likelihood, the total variability of NOIRF may also depend on parameters out of this list, such as atmospheric tidal effects \citep{nischal2019solar,oberheide2013impact}. 
\begin{figure}[h!]
\noindent\includegraphics[width=\textwidth]{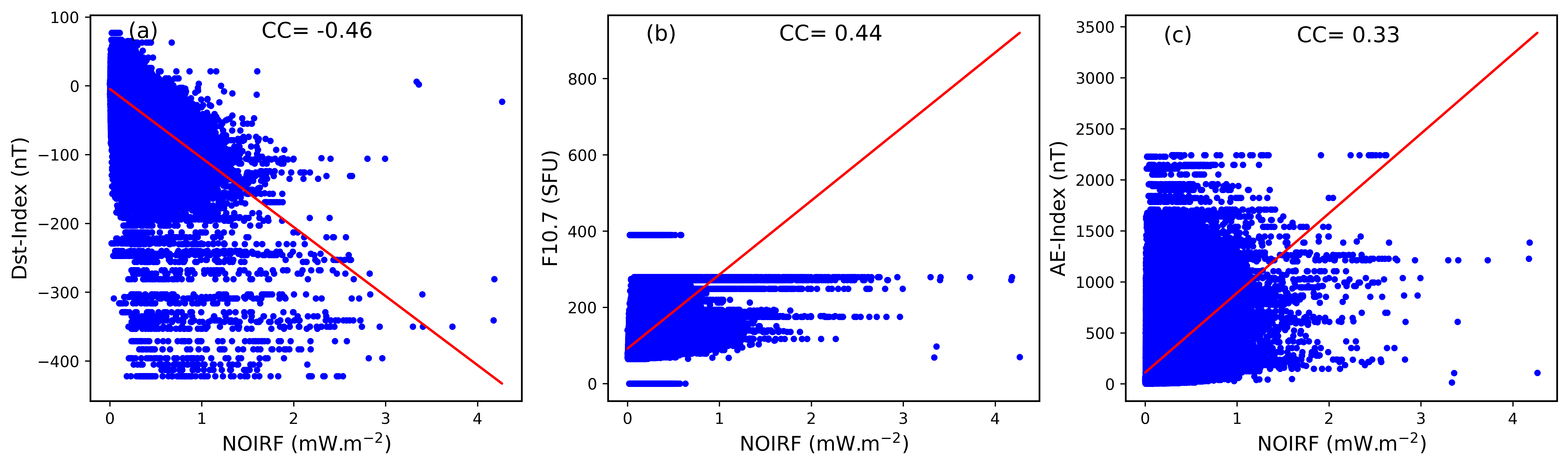}
\caption{Correlation of measured NOIRF by SABER with (a) Dst index, (b) F10.7, and (c) AE-index.}
\label{pngfiguresample}
\end{figure}
Due to a lack of data, such parameters have been excluded from our list. Some of the investigated parameters/features mentioned in Table 1 have a significant correlation with the target parameter (NOIRF) and are the primary cause of its variation.
Among the various features, the Dst index has been found to be strongly anti-correlated (-0.46) to the NOIRF, and the maximum correlation is (0.44), which is obtained with the F10.7 flux. The AE index known to be directly related to the NOIRF in the thermosphere is positively correlated with a factor of 0.33. The correlation of NOIRF with these parameters based on the SABER observations (2002-2017) is shown in Figure 2.

\section{Algorithm and implementation}
In the present work, we have used regression algorithms based on supervised learning, a subset of machine learning and artificial intelligence approach that can be used for building highly accurate machine learning models \citep{mahesh2020machine}. Based on our tasks and data structure, we used supervised ensemble machine learning algorithms like Decision Tree, Extreme Gradient Boosting (XGBoost) \citep{chen2016xgboost}, and Random Forest \citep{ho1998random,breiman2001random} to train our model. These three ML models have been used, and their performance is evaluated based on the actual data. The model with the highest R-value has been chosen for predictive modeling. The R-value is a statistical measure that represents the goodness of fit of a regression model. It is also known as the linear correlation coefficient, which has a value between -1 and 1. A higher R-value represents the more accurate model. R-value equals 1 when the model perfectly fits the data, and there is no difference between the predicted and actual values. The Decision Tree algorithm is a supervised non-parametric learning method that can be used for regression as well as classification tasks \citep{pathak2018assessment}. It requires little data preparation and can handle multi-output problems for both numerical and categorical data types. It is also possible to validate this method using statistical tests, which makes the model reliable.
The Random Forest (RF) technique \citep{ho1998random,breiman2001random} is a combination of several decision trees. To obtain a minimal variance, RF takes an average of numerous decision trees. In this study, an RF regressor with 150 decision trees is used for training the model. The third machine learning algorithm used in this study is known as XGBoost \citep{chen2016xgboost}. It is based on decision trees and is an ensemble method developed to prevent overfitting and for effective parallel processing. It uses the gradient descent algorithm to minimize the loss and thus creates a model with good accuracy. Here, we used XGBoost at a learning rate of 0.3 with 300 estimators and a maximum depth of 15. We apply these three algorithms to the same training and testing datasets. The comparison and performance results for the three algorithms with their R-values are shown in Table 2. The Higher R-value indicates the higher reliability and accuracy of the respective model. 
\begin{table*}[h]
\centering
\caption{R-value of various ML algorithms for training and test dataset.}
\begin{tabular}{|c|c|c|}
\hline
ML/AI Algorithm &   \multicolumn{2}{|c|}{R-value}  \\

    & Training  & Testing\\
    \hline
    Decision Tree & 0.99 & 0.93\\
    XGBoost &	0.99&	0.95\\
    Random Forest&	0.99&	0.96\\
    \hline
\end{tabular}
\end{table*}

All the above-mentioned work has been implemented in Python programming language \citep{van1995python} using Scikit-learn\citep{pedregosa2011scikit} and Keras \citep{chollet2015keras} libraries.

\section{Model performance:}
The objective of the present work is to develop an ML-based model to estimate and predict the value of NOIRF using the SABER data for a period of 16 years (2002-2017). For this purpose, we have merged the SABER and OMNIWeb data and applied various machine learning algorithms, as discussed in section 2. The entire dataset has been split into two subsets. One subset from 2002-2014 is used for the model training and testing. The other subset from 2015-2017 is used to evaluate the performance of the model.  It is very important to note that the second subset of the data is never exposed to the model. This is the key feature of this study, and this approach helps develop a functional predictive model with the algorithm chosen with the given efficiency. 
To train and test the model we used the main dataset 2002-2014 and split this again into two subsets known as the training set and testing set. The subsets for training and testing contain 90\% and 10\% of data, respectively, from the main dataset. The subset with 90\% data is used for training, and the remaining 10\% subset is used to test the model. The data for training and testing subsets were selected randomly from the main dataset (2002-2014) without any bias to the prevalent geophysical conditions. This allows the ML model to develop finer connections between the features and the target parameter.  For all ML algorithms discussed earlier, the same training dataset has been used so that the accuracy of the model can be compared on an equal footing. Amongst the three algorithms, Random Forest produces a better accuracy in terms of R-value. Hence, this algorithm is further used to develop the model, make predictions, and compare the predictions with actual measurements. 
The NOIRF obtained from SABER measurements against the input features, such as latitude, longitude, time, etc, is compared with the ML model (NOEMLM) derived IRF for the same set of features. The correlation between the predicted and the actual value of log$_{10}$(NOIRF) for training and testing datasets is shown in Figure 3.
\begin{figure}[h]
\noindent\includegraphics[width=\textwidth]{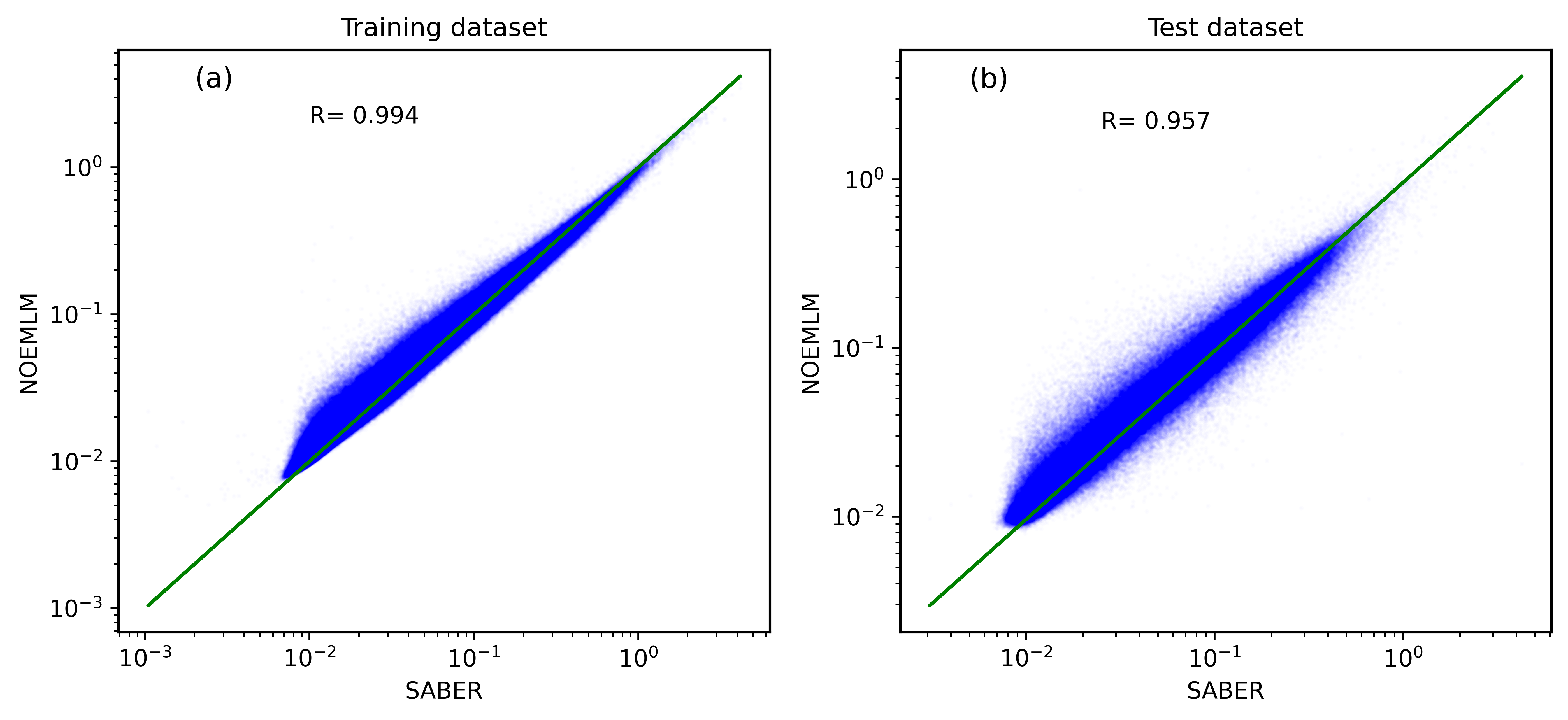}
\caption{The correlation between NOEMLM predicted and SABER observed NOIRF for the (a) training and (b) test datasets. The unit of NOIRF is mW.m$^{-2}$.}
\label{pngfiguresample}
\end{figure}

The green line in Figure 3 corresponds to the absolute agreement between the predicted NOIRF by NOEMLM and the measurements of SABER. This line is the best-fit line, for which we have zero deviation from the true values of NOIRF. The deviation from the green line measures the difference between the model prediction vs the actual value of NOIRF. The accuracy of the model for the training and testing set is presented by the R-value. A higher value of R means a higher accuracy of prediction. 
It can be seen from Figures 3(a) and 3(b) that the slope of log$_{10}$(NOIRF) stands at 0.994 for the training dataset and 0.957 for the testing dataset. The modeled NOIRF is correlated positively with a coefficient of 0.988 and 0.918 for the training and testing datasets, respectively. These coefficients indicate that the ML model is able to accurately predict 98.8\% of the variability in NOIRF within the training dataset and 91.8\% in the testing dataset. The absolute value of the correlation coefficient indicates the better-performing ability of the ML model for the given dataset. To quantify the prediction accuracy of the current ML model, the root mean square error (RMSE) has been calculated for the predicted NOIRF against SABER observations during 2015-2017. The RMSE is calculated as follows.
\vspace{2mm}
\begin{center}
     Root Mean Square Error (RMSE) = $\sqrt{\frac{(\Sigma^N_{i=1}(y-y')^2}N}$

      Root Mean Square percentage Error (RMSE\%) = $\frac{1}{N}\sqrt{\frac{(\Sigma^N_{i=1}(y-y')^2}y}\times 100$
 
\end{center}{}
Where,  i = variable

		y =  SABER measured actual values of NOIRF
		
		y’ = NOEMLM predicted value of NOIRF
		
		N = Total number of data points

With respect to the testing dataset, An RMSE of 0.0468 mW.m$^{-2}$ has been obtained. The low magnitude of RMSE indicates the better prediction ability of NOEMLM, which also complements the earlier discussion based on correlation coefficients. A set of all input features (listed in Table 1) for a single event in the observational dataset (between 2015 and 2017) is defined as an Index. The NOIRF measured against this Index by SABER and the NOEMLM prediction for the same Index is shown in Figure 4. 
\begin{figure}[h]
\noindent\includegraphics[width=\textwidth]{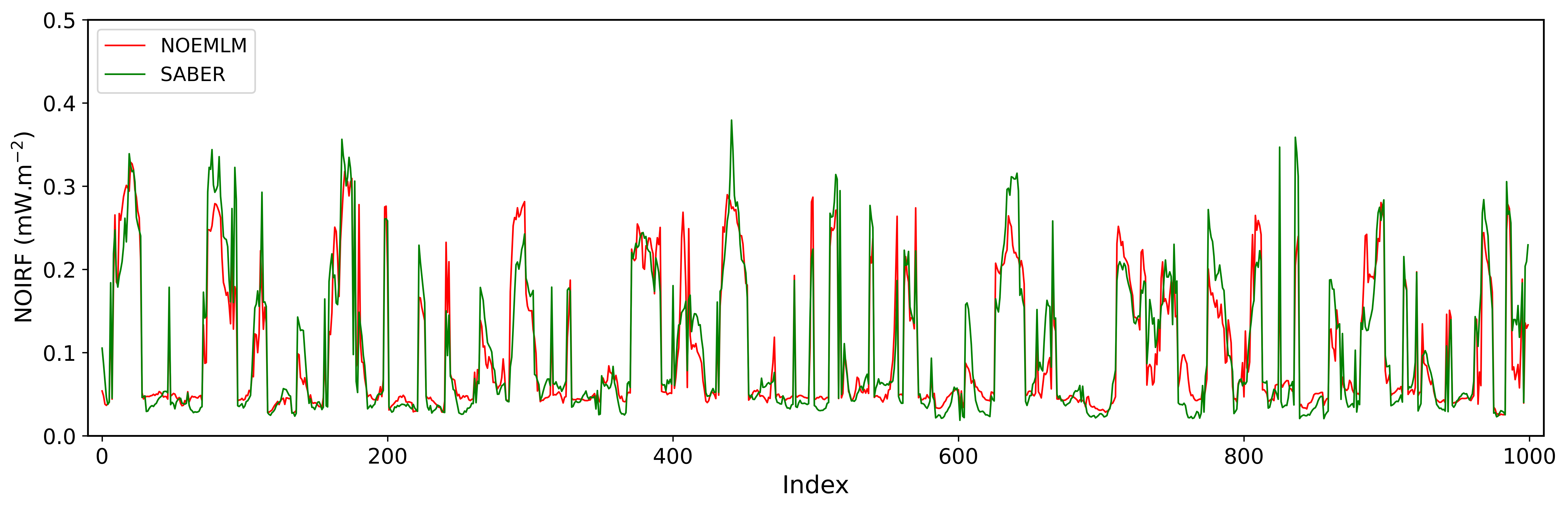}
\caption{The NOIRF values predicted by NOEMLM in comparison to the observations of SABER. The test samples were selected randomly from the dataset (2015-2017)}
\label{pngfiguresample}
\end{figure}

It can be seen from the figure that the model predictions match the observational dataset quite well at all Index values. The daily averaged NOIRF predicted by the NOEMLM for the observational period of three years (2015-2017), along with the actual SABER measurements, are shown in Figure 5. 

\begin{figure}[h]
\noindent\includegraphics[width=\textwidth]{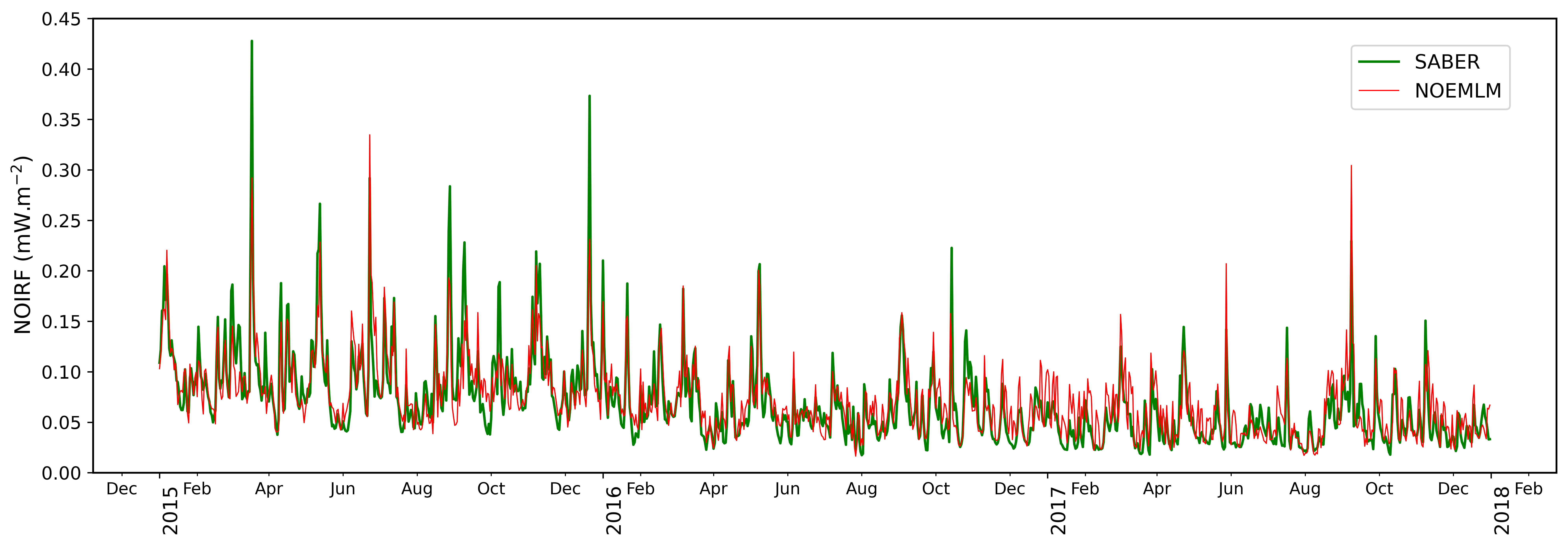}
\caption{Daily averaged NOIRF as predicted by NOEMLM, in comparison to SABER observations.}
\label{pngfiguresample}
\end{figure}

As established from the discussion in the earlier sections, it is evident that the NOEMLM is able to make finer connections between the target variables and input features and is also able to weigh in the subtle relations among the features for the observed fluctuation in NOIRF. The period 2015-2017 corresponds to the descending phase of solar cycle 24, which has witnessed a few very intense geomagnetic storms. The large-scale geomagnetic fluctuations affect various input features simultaneously, and hence, it is a challenging task to estimate the combined effect of these events on the target variable. It would be a worthwhile effort to test the NOEMLM for the understanding and estimation of the geomagnetic forcing on the NOIRF. The testing of the ML model for selected intense space weather fluctuations during the period 2015-2017 is discussed in the sections ahead.

\subsection{NOIRF prediction during space weather events:}
For the purpose of testing the NOEMLM’s prediction ability of NOIRF during intense space weather events, two severe geomagnetic storms, which occurred during 22-25 June 2015 (Storm 1) and 7-9 September 2017 (Storm 2), have been chosen. A geoeffective coronal mass ejection associated with an M-class flare on the Sun resulted in a moderately severe G4-class geomagnetic storm on 22 June 2015 \citep{augusto20182015}. This geomagnetic storm was the second most intense storm in the solar cycle 24. The main phase of the storm lasted for two days, followed by the recovery phase. The impact of CME on the magnetosphere and subsequent changes in the earth’s magnetic field is quantified by the Dst index \citep{gonzalez1994geomagnetic}. It essentially indicates the decrease in the horizontal component of the earth’s magnetic field during the time of enhanced ring current result of gradient and curvature drifts of highly energetic electrons and protons injected in the inner magnetosphere due to multiple magnetic reconnections into the nightside during geomagnetic storms. The Auroral Electrojet Index (AE) is a quantitative measure of auroral zone magnetic activity produced by the enhanced ionospheric currents flowing within and below the auroral oval \citep{davis1966auroral}.

\begin{figure}
\noindent\includegraphics[width=\textwidth]{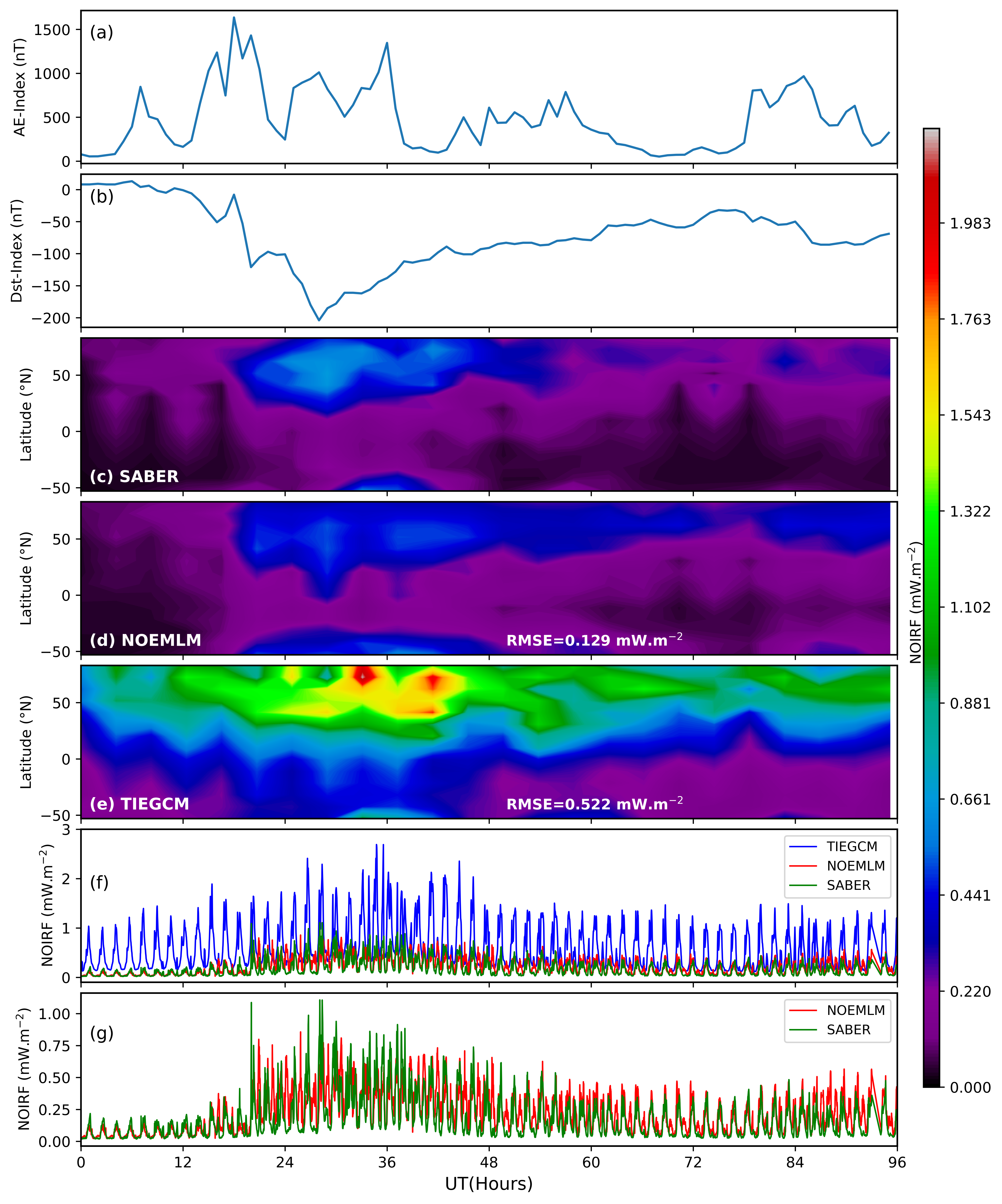}
\caption{Variation of (a) AE-index, (b) Dst index, (c) SABER observed global NOIRF, (d) NOEMLM predicted global NOIRF, (e) TIEGCM calculated NOIRF, and (f, g) time series evolution of NOIRF during Storm 1 (22-25 June 2015).}
\label{pngfiguresample}
\end{figure}

The variation in Dst and AE-index during Storm 1 (22-25 June 2015) is shown in Figure 6 (a \& b). As the storm time energy enters the Earth’s atmosphere at the polar latitudes, there is a substantial enhancement in the total NOIRF, which can be seen from SABER measurements (Figure 6(c)). The enhancement is centered at the high latitudes and is spread towards the mid and low latitudes due to the transport of depleted O/N$_2$ by means of strong meridional wind at high velocities \citep{zhang2004n2}. The enhancement in NOIRF persists over several days during the recovery phase before it reaches normal values towards 25 June 2015. The enhancement in the southern polar region cannot be clearly seen as the SABER instrument was in the north viewing mode at that time.

In order to evaluate the performance of NOEMLM, it has been implemented to predict the storm time variation of NOIRF during Storm 1. The predicted NOIRF for Storm 1 is shown in Figure 6(d). Although the observational period of 2015-2017 is completely new to the NOEMLM, the figure shows that the model captures the variability of NOIRF with respect to the Dst and AE- index very well. The penetration of NO cooling emissions during Storm 1 is larger in the northern hemisphere, as expected due to the background summer to winter wind in addition to the joule heating induced meridional wind; this wind brings the nitrogen (N$_2$) and NO abundant air towards lower latitude, which ultimately results in higher NOIRF in low to mid-latitude of the summer hemisphere in comparison to the winter hemisphere. It can be seen from Figure 6 (d) that NOEMLM is able to capture the latitudinal variability of NOIRF for quiet as well as storm time which can be considered a great achievement for NOEMLM to be able to predict this effect. 

The NOEMLM predictions during Storm 1 are compared with the existing Thermosphere-Ionosphere- Electrodynamics General Circulation Model (TIEGCM) shown in Figure 6 (e). By observing Figure 6 (e), it is clearly seen that the TIEGCM overestimates the NOIRF compared to SABER, and its predictions are far ahead of the actual measurements. For TIEGCM, an RMSE of 0.522 mW.m$^{-2}$ is calculated, which shows a higher deviation from the actual values of NOIRF. The comparison results show that the NOEMLM predictions have much better accuracy than the TIEGCM and hence can be considered a better tool for predicting the space weather fluctuations induced in the radiative cooling emissions. The time series comparison of NOIRF predictions by NOEMLM and TIEGCM compared to SABER is shown in Figure 6 (f). For better visualization, the comparison results of NOEMLM prediction and SABER with the time are shown in Figure 6 (g).

The NOEMLM effectively captures the variability and the enhancement of NOIRF very well, but it can also be seen that the model predictions are slightly off in comparison to the SABER observations in terms of numerical values. To quantify this, an RMSE of 0.129 mW.m$^{-2}$ calculated during Storm 1 assures the reliability of the NOEMLM to predict the storm-time variation of NOIRF. This low RMSE also suggests that the model can further be used to know the possible effect of geomagnetic storms on the Earth’s atmosphere in the context of radiative cooling processes. The differences in measured and predicted values may be attributed to several factors; the lack of a complete set of dependent variables against the NOIRF in the measurement dataset could be the major one. The day/night asymmetry in terms of SABER coverage could also bring some uncertainty to the agreement. Including these parameters may further enhance the model’s prediction accuracy. 

During the descending phase of solar cycle 24, a coronal mass ejection associated with an X-9 flare resulted in a G3 class geomagnetic storm on 7 September 2017. The peculiarity of this storm is that the total duration of the storm is associated with two depletion phases with minimum Dst indices of -142 and -122 nT. The entire storm duration has two main phases with high intensity separated by nearly 14 hours \citep{sun2022responses}. The variability of NOIRF during double storms has been observed to be quite different with respect to a normal geomagnetic storm \citep{bharti2018storm, ranjan2023aspects}. Thus, this storm provides a unique case for understanding the NOIRF variability when there is not sufficient time for the entire atmosphere to recover from a storm before another intense geomagnetic storm occurs.

\begin{figure}
\noindent\includegraphics[width=\textwidth]{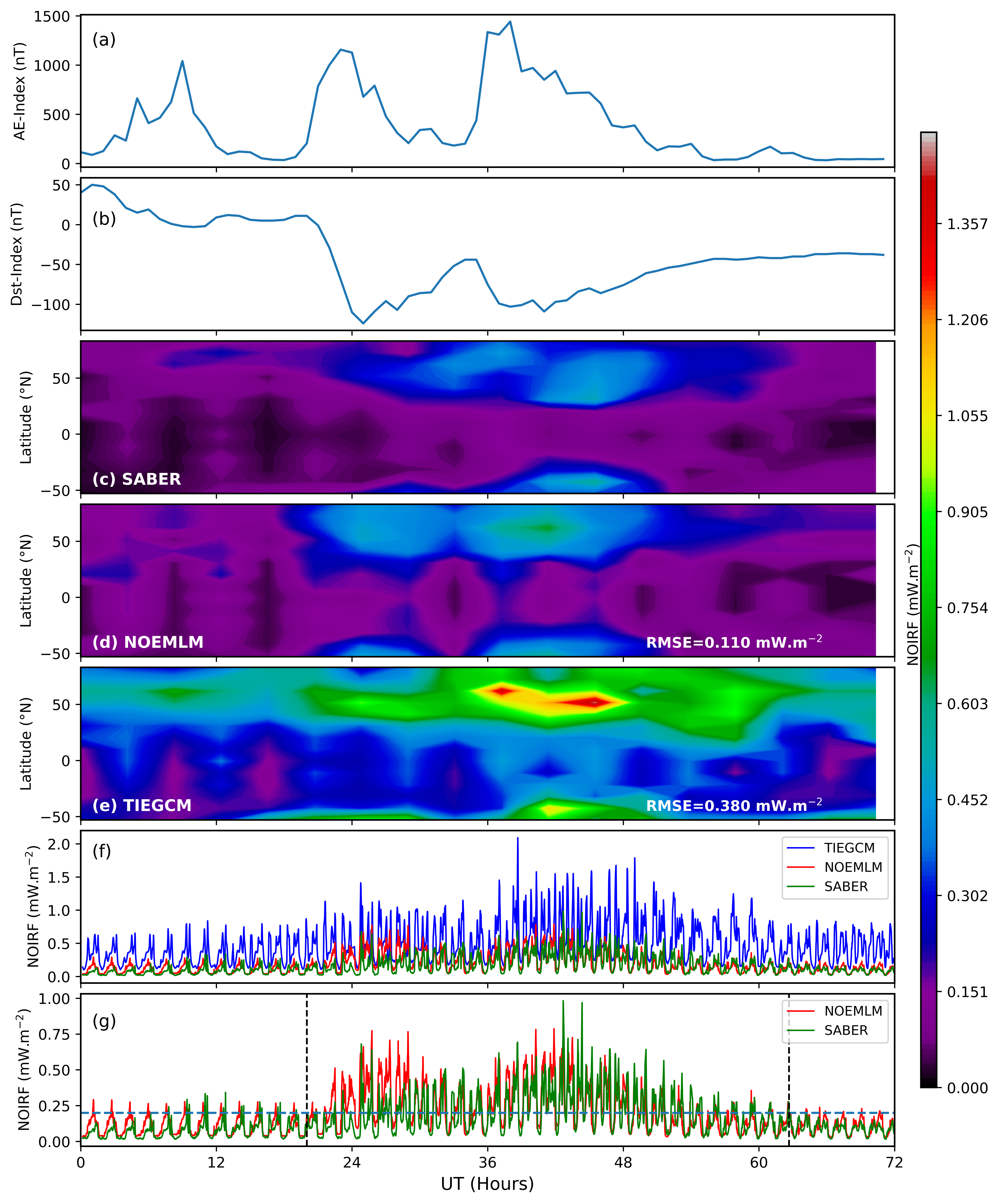}
\caption{Variation of (a) AE-index, (b) Dst index, (c) SABER observed global NOIRF, (d) NOEMLM predicted global NOIRF, (e) TIEGCM calculated NOIRF, and (f,g) time series evolution of NOIRF during Storm 2 (7-9 September 2017).}
\label{pngfiguresample}
\end{figure}

The storm time variation of the AE index and Dst index is shown in Figure 7 (a) and (b), respectively. The SABER retrieved NOIRF obtained using the procedure explained in earlier sections is shown in Figure 7 (c). The effect of a double storm on radiative cooling can be clearly seen in Figure 7 (c). It can be noticed that the enhancement in the NOIRF is more pronounced during the second main phase (14:00 UT on 8 September 2017) in comparison to the first main phase (01:00 UT on 8 September 2017), although the first main phase is more intense in magnitude \citep{sun2022responses}. This can be due to the pre-conditioning of the upper atmosphere by the first main phase. More importantly, the equatorward spread of enhanced NOIRF can be very clearly seen. The NOEMLM is used to predict the global variation in the NOIRF during the entire duration of the double storm. The predicted value of NOIRF is completely based on the training dataset. It can be seen that the predicted NOIRF with an RMSE of 0.110 mW.m$^{-2}$ agrees very closely with the actual SABER measurements. The basic nature of the fluctuation in NOIRF during the double storm is successfully captured by the NOEMLM. 
As it was seen in the correlation investigation of NOIRF emissions with various solar activity indices, the NOIRF has a larger negative correlation with Dst (-0.46) and a positive correlation of 0.33 with the AE index. However, NOEMLM is able to predict comparatively higher NOIRF during the second phase of Storm 2 (Figure 7 (d)), where the AE index is higher and the Dst index is lower compared to the first main phase. The same results are also observed by SABER during the second phase of Storm 2, as seen in Figure 7 (c). The AE index is a proxy of joule heating during geomagnetic storm \citep{ahn1983joule,baumjohann1984hemispherical}. During the second main phase of Storm 2, the higher AE index represents the higher amount of joule heating, which results in the higher NOIRF as observed by SABER and predicted by NOEMLM. This also shows the accuracy of NOEMLM in effectively capturing and predicting variation in NOIRF in extreme space weather events and quiet time. 
The NOEMLM’s predictions during Storm 2 (7-9 September 2017) were also compared with the existing chemistry-based model TIEGCM, and comparison results are shown in Figure 7. By observing Figure 7 (e), it is clearly visible that the TIEGCM predictions significantly overestimated the actual SABER observations and NOEMLM predictions. The root mean square error (RMSE) was calculated to quantify the performance of both models. The calculated RMSE values for NOEMLM and TIEGCM are 0.110 mW.m$^{-2}$ and 0.380 mW.m$^{-2}$, respectively. These results indicate that NOEMLM demonstrates better capability in predicting NOIRF as compared to TIEGCM. This also suggests that the machine learning (ML) technique presented in this paper is much more effective in capturing complex atmospheric dynamics and processes compared to traditional chemistry-based climatological models. Figure 7 (f) shows the time-series variation of NOIRF during the entire duration of the storm. For better visualization of the temporal variation in NOIRF, the comparison results of NOEMLM prediction with SABER observation are shown in Figure 7 (g).

However, the RMSE values calculated for both storms represent the all-over performance of the model at all latitudes during the geomagnetic storms. During the geomagnetic storm, the higher latitudes show complex behavior compared to lower, so it would be better to test the ability of NOEMLM to capture NOIRF variation over higher as well as lower latitudes. To do this, we consider a range of latitudes listed in Table 3 and test the prediction accuracy of NOEMLM compared to the SABER observations. To quantify the performance of NOEMLM at each latitude range, we calculate the error in terms of RMSE for both storms (Storm 1 and Storm 2) listed in Table 3. It can be clearly seen from Table 3 that the accuracy of the model prediction varies with the latitude range. For lower latitudes, the predictions are closer to the measurements, and for mid to higher latitudes, the accuracy becomes lower. As it is known, the nature of physical and electrodynamical interactions that occur over the high latitude thermosphere are highly complex, and there is a need to identify and include other drivers to achieve better prediction accuracy.
\begin{table*}[h]
\centering
\caption{The error in prediction accuracy of NOEMLM in terms of RMSE for various latitude ranges.}
\begin{tabular}{|c|c|c|}
\hline
Latitude (in Degree) &   \multicolumn{2}{|c|}{RMSE (in \%)}  \\
& Storm 1 & Storm 2 \\
\hline
70-83 & 12.60 &12.45  \\
 60-70 &14.33	&	14.56\\
 50-60&13.79	&12.58	\\
-40 to -50 \& 40 to 50 &17.89	&	13.07	\\
-30 to -40 \& 30 to 40 	& 12.68&	12.55\\
-20 to -30 \& 20 to 30 & 09.14&		09.40\\
-10 to -20 \& 10 to 20 & 06.98&	07.12	\\
-10 to 10 &07.59	&	06.11\\
\hline
\end{tabular}
\end{table*}

The time series variation from Figures 6 (g) \& 7 (g) shows that the NOEMLM has a good agreement to capture the variation with time and the progression of the storm. The predicted value of NOIRF by NOEMLM corresponds to the same time for each value observed by SABER. To test the ability of NOEMLM to capture the temporal variation of NOIRF during the initial, main, and recovery phases of geomagnetic storms, we consider Storm 2. The recovery phase of Storm 1 is more than ten days, and it overlaps with other geomagnetic activity during its recovery phase. Due to this, Storm 1 could not recover completely and could not achieve its pre-storm value of NOIRF. However, in the case of Storm 2, having a time period of nearly three days, it recovered within this window of three days and achieved its pre-storm value of NOIRF about 0.158 mW.m$^{-2}$ (less than 0.2 mW.m$^{-2}$, shown by horizontal line) after 42.66 hours as marked by vertical line nearly 62.66 UT in Figure 7 (g). The predicted value of NOIRF at the same time by NOEMLM is 0.186 mW.m$^{-2}$, and the TIEGCM estimated value is 0.504 mW.m$^{-2}$. It can be concluded that the NOEMLM predicted value is closer to the actual measurement of SABER. The findings indicate that the NOEMLM demonstrates good latitudinal and temporal predictive capability for NOIRF during various geophysical conditions.

\section{Discussion and Summary}
This research examines how space weather primarily affects the Earth’s upper atmosphere in terms of radiative cooling by NO. The variation of NO during space weather events has been attempted in many earlier studies. The observed variation of several chemical species and their overall effect on radiative cooling is mentioned in the introduction. The central objective of this study is to develop a working machine-learning model that can capture the climatological observations of NOIRF by SABER and produce good accuracy in predicting the fluctuations. This study, however, does not delve into the chemistry and physics of a specific species. Instead, it describes the behavior of the upper atmosphere during quiet and storm times based on the NOIRF as an indicator of higher solar electron and proton precipitation in the upper atmosphere of Earth.
In this study, random forest machine learning algorithms are used to develop a predictive model for the NOIRF. The choice of algorithm is based on the model’s performance on the test dataset. The model is trained with thirteen years of SABER observations of NOIRF in the altitude range of 115-250 km of Earth’s atmosphere. The satellite data is successfully merged with other magnetic indices and solar proxies, which are representative of the various geophysical conditions prevalent during the observation period of SABER. The objective of the model is to successfully predict the variability of NOIRF with respect to latitude, longitude, and time during various space weather events and also with respect to the progression of the solar cycle. The NOIRF can be considered an essential tracer to understand the energetics and structure of the upper atmosphere, mainly the thermosphere during space weather fluctuations.
The NOEMLM tested with actual measurements and is able to capture 91.8\% accurate variations in NOIRF for the testing dataset. This indicates that the NOEMLM has successfully been able to capture the very nature of fluctuations in NOIRF with respect to several other independent variables. The model once validated with good R-values has been further used to predict NOIRF during intense geomagnetic storms which occurred in the years 2015 and 2017. The global predicted NOIRF is compared with the actual measurements, and it has been found that the model demonstrates very good prediction accuracy. A few very important connections between space weather and radiative flux are very well represented in the model output.  The model performed very well during extreme space weather events, specifically Storm 1 and Storm 2. For both geomagnetic storms, the NOIRF pattern and upper atmospheric dynamics were very well captured by the NOEMLM with the time and latitudes.

For both measured and predicted NOIRF, the intensity is found to have an inverse relationship with the Dst index, while the latitudinal penetration strongly correlates with the AE index during geomagnetic storms. This demonstrates that Dst and AE indices are useful parameters for understanding atmospheric behavior during such events, and the model successfully accounts for their effects on the upper atmosphere of Earth. The model indicates that for higher AE indices, there is greater penetration towards lower latitudes, which can be explained by the general physics of the AE index, whereby stronger AE indices generate stronger currents in the ionosphere that interact with the Earth's magnetic field at lower latitudes.
During geomagnetic storms, the density, temperature, and composition of the thermosphere undergo rapid and non-uniform changes, leading to variable densities and temperature distributions. This may be the major cause for the overestimation of NOIRF by TIEGCM. These changes are primarily caused by solar radiation, magnetic fields, and particle precipitation. To proxy these changes in NOEMLM, the F10.7, Dst, and AE indices have been used to estimate changes in temperature, density, and chemical composition. The learning algorithm calculates the correlation factor for each input using training data and is optimized to estimate the NOIRF based on these controlling factors (Dst, AE, F10.7, etc.).
A small deviation in the predicted value of NOIRF by NOEMLM could be due to the absence of concurrent measurements of real-time data for various controlling factors (AE, Dst, etc.), which were considered uniform for a small-time range due to lack of real-time data. This could be a significant reason for the deviation of predicted values compared with the SABER, especially during the geomagnetic storm. Additionally, day/night asymmetry in SABER coverage may also contribute to uncertainty in the agreement.

However, in comparison to the existing TIEGCM model, NOEMLM has excellent performance, especially during storm time, when the behavior of the atmosphere became very complex. From Figure 6 (d, e) \& 7 (d, e) it is clear that the predictions of TIEGCM are much higher than the actual measurements.
The findings of this study suggest that utilizing geomagnetic and space weather indices can serve as superior parameters for studying the upper atmosphere, as compared to focusing on specific species having complex chemical processes and associated uncertainties in constituents. Additionally, machine learning techniques can effectively carry out this analysis with greater ease than traditional chemical studies.
This study presents a working machine learning model which can be used as an effective tool in predicting space weather influence on the thermosphere. This work also establishes the ML methodology as an effective tool for understanding the complex and dynamic behavior of Earth’s atmosphere when the chemistry of the atmosphere changes dramatically.

\section{Acknowledgments}
We acknowledge the \href{http://saber.gats-inc.com/data.php}{SABER},  \href{https://omniweb.gsfc.nasa.gov/}{OMNIWeb}, \href{https://lasp.colorado.edu/snoe/}{SNOE}, and  \href{https://ccmc.gsfc.nasa.gov/}{Community Coordinated Modeling Center (CCMC)} for providing all necessary data for developing this NOIRF model. The OMNIWeb data is available at \url{https://omniweb.gsfc.nasa.gov/form/dx1.html}, and SABER data is available at: \url{http://saber.gats-inc.com/data.php}. The SNOE data can be accessed from \url{https://lasp.colorado.edu/snoe/data/download-data/}.
The TIEGCM simulation results have been provided by CCMC at Goddard Space Flight Center through their publicly available simulation services {\url{https://ccmc.gsfc.nasa.gov}. One of the authors D.Nailwal thanks the Ministry of Human Resource Development (MHRD), Government of India for financial support as a graduate assistantship.

\bibliography{bibliography}{}
\bibliographystyle{aasjournal}

\end{document}